\documentstyle[12pt]{article}
\newcommand{\norm}[1]{{\protect\normalsize{#1}}}
\newcommand{\LAP}
{{\small E}\norm{N}{\large S}{\Large L}{\large A}\norm{P}{\small P}}

\newcommand{\A}{\alpha}
\newcommand{\B}{\beta}

\newcommand{\G}{\gamma}

\newcommand{\be}{\begin{equation}}
\newcommand{\ee}{\end{equation}}
\newcommand{\bea}{\begin{eqnarray}}
\newcommand{\ena}{\end{eqnarray}}
\newcommand{\beano}{\begin{eqnarray*}}
\newcommand{\enano}{\end{eqnarray*}}
\newcommand{\nn}{\nonumber \\ }
\newcommand{\sect}[1]{\setcounter{equation}{0}\section{#1}}
\newcommand{\vs}[1]{\rule[- #1 mm]{0mm}{#1 mm}}
\newcommand{\hs}[1]{\hspace{#1 mm}}

\newcommand{\half}{\frac{1}{2}}

\newcommand{\ca}{\mbox{$\cal{A}$}}
\newcommand{\cb}{\mbox{$\cal{B}$}}

\newcommand{{\cg}}{\mbox{$\cal{G}$}}
\newcommand{\ch}{\mbox{$\cal{H}$}}

\newcommand{\cn}{\mbox{$\cal{N}$}}

\newcommand{\ct}{\mbox{$\cal{T}$}}
\newcommand{\cw}{\mbox{$\cal{W}$}}
\newcommand{\cy}{\mbox{$\cal{Y}$}}

\newcommand{{\sg}}{\mbox{$\cal{SG}$}}
\newcommand{\vph}{\varphi}

\newcommand{\prt}{\partial}
\newcommand{\pa}{\partial}
\newcommand{\eps}{\epsilon}
\newcommand{\bal}{{\bar{\alpha}}}

\newcommand{\wh}[1]{\widehat{#1}}
\newcommand{\mb}[1]{\hs{5}\mbox{#1}\hs{5}}

\newcommand{\su}{\mbox{${s\ell(2)}$}}
\newcommand{\sln}{\mbox{${s\ell(n)}$}}

\newtheorem{lem}{Lemma}

\newcommand{\C}{\mbox{\hspace{.0em}\rule{0.042em}{.674em}\hspace{-.642em}
\rm C$\,$}}

\newcommand{\1}{\mbox{\hspace{.0em}\rule{0.042em}{.7em}\hspace{-.524em}
{\large 1}$\!$}}
\newcommand{\itN}{I\hspace{-0.27em}N}

\newcommand{\NP}[1]{Nucl.\ Phys.\ {\bf #1}}
\newcommand{\AP}[1]{Ann.\ Phys.\ {\bf #1}}
\newcommand{\PL}[1]{Phys.\ Lett.\ {\bf #1}}

\newcommand{\CMP}[1]{Comm.\ Math.\ Phys.\ {\bf #1}}

\newcommand{\IJMP}[1]{Int. Journ.\ Mod.\ Phys.\ {\bf #1}}

\def\ord#1#2{{#2\over (z-w)^{#1}}}
\def\ordo#1{{#1\over z-w}}
\def\ope#1#2{#1(z)#2(w)}
\newcommand{\hf}{\frac{1}{2}}
\newcommand{\th}{\frac{1}{3}}
\newcommand{\fo}{\frac{1}{4}}
\newcommand{\ei}{\frac{1}{8}}
\newcommand{\thf}{\frac{3}{2}}

\begin{document}
\renewcommand{\thefootnote}{\fnsymbol{footnote}}
\newpage
\pagestyle{empty}
\setcounter{page}{0}


\hbox to \hsize{\vbox{\hsize 8 em
{\bf \centerline{Groupe d'Annecy}
\ \par
\centerline{Laboratoire}
\centerline{d'Annecy-le-Vieux de}
\centerline{Physique des Particules}}}
\hfill
\hfill
\vbox{\hsize 7 em
{\bf \centerline{Groupe de Lyon}
\ \par
\centerline{Ecole Normale}
\centerline{Sup\'erieure de Lyon}}}}
\hrule height.42mm

\vs{10}
\begin{center}

{\LARGE {\bf Linearization of \cw\ algebras\\ and \cw\ superalgebras}}\\[1cm]

\vs{2}

{\large J.O. Madsen\footnote{e-mail: madsen@lapphp8.in2p3.fr}, and
E. Ragoucy\footnote{e-mail: ragoucy@lapp.in2p3.fr}}\\[4mm]

{\em Laboratoire de Physique Th\'eorique }\LAP\footnote{URA 14-36
du CNRS, associ\'ee \`a l'Ecole Normale Sup\'erieure de Lyon
et \`a l'Universit\'e de Savoie}\\

Groupe d'Annecy, Chemin de Bellevue BP 110\\
F-74941 Annecy-le-Vieux Cedex, France.

\end{center}
\vfill

\centerline{ {\bf Abstract}}

\indent

In a recent paper, the authors have shown that the secondary reduction of
\cw-algebras provides a
natural framework for the linearization of \cw-algebras. In particular, it
allows in a very simple way
the calculation of the linear algebra $\cw(\cg,\ch)_{\geq0}$ associated to
a wide class of \cw(\cg,\ch)
algebras, as well as the  expression of the $W$ generators of \cw(\cg,\ch)
in terms of the generators
of $\cw(\cg,\ch)_{\geq0}$.

In this paper, we present the extension of the above technique to
\cw-superalgebras, i.e. \cw-algebras
containing fermions and bosons of arbitrary (positive) spins. To be
self-contained the paper recall
the linearization of \cw-algebras. We include also examples such as
the linearization of $\cw_n$
algebras; $\cw(s\ell(3|1),s\ell(3))$ and $\cw(osp(1|4),sp(4))\equiv
\cw B_2$ superalgebras.
\vfill
\rightline{hep-th/9510061}
\rightline{\LAP-A-520/95}
\rightline{October 1995}

\newpage
\pagestyle{plain}
\setcounter{footnote}{0}
\renewcommand{\thefootnote}{\arabic{footnote}}
\sect{Introduction}

\indent

Recently there has been much interest in the linearization of \cw-algebras.
This
linearization consists in adding a number of fields to a \cw-algebra, in such
a way that the resulting algebra is equivalent, by a nonlinear,
{\em invertible}
basis-transformation, to a linear algebra. Such a linearization was first
proposed for $\cw_3$ and $\cw_3^2$ by Krivonos and Sorin in \cite{KrSo}, and
for $\cw_{(2,4)}$ and $\cw\cb_2$ in \cite{BeKrSo}. In \cite{JOE} the present
authors proposed a general method for the linearization of a large class
of quantum \cw~algebras, using the method of secondary quantum hamiltonian
reduction. The advantage of such a technique is to give a general framework
for the linearization and to
use only  fields of positive spin.
In \cite{KrSo2}, a different method of linearization for superconformal
algebras (i.e.
algebras with one spin 2 and spin $\frac{3}{2}$ and spin 1 fields)
was given. These algebras are not linearizable by our techniques, since their
linearization uses fields of spin $-\frac{1}{2}$.
Note that a conjecture about the linearization of the
other \cw-algebras, using superconformal algebras, spin $(-\frac{1}{2})$
fields and secondary
reductions was also given in the same paper.

The purpose of the present paper is to explain in more detail the
linearization proposed in \cite{JOE}, to show how to apply the method to
classical \cw~algebras, and to extend the method to \cw~algebras that
includes fermionic fields, i.e. \cw~algebras that are obtained by the
hamiltonian reduction of Lie superalgebras. In particular, we want to
stress that our procedure
indeed provides explicit realizations for the linearization of \cw-algebras
and
superalgebras.

\indent

In a first section, we recall the linearization of \cw(\cg,\ch) algebras and
treat explicitly the case of
$\cw_n\equiv\cw(\sln)$ algebras. Then, in the second section, we show that
the same procedure can
be applied to superalgebras. Section 3 is devoted to two examples,
one based on
$s\ell(3|1)$, and the other on $osp(1|4)$. Finally, the last section conclude
on open questions. We have also added a technical appendix where the most
important parts of the
conjecture given in \cite{JOE} is proven.

\indent

For explicit calculations, we have used the Mathematica package of
K. Thielemans \cite{OPE}.

\sect{Linearization of \cw-algebras}
\subsection{Secondary reductions}

\indent

To be self-contained, we briefly recall the framework of secondary
reductions. For details, see the
original papers on secondary reductions\cite{Wgd,JOE}, for a review
on \cw-algebras, see
\cite{BoSc}, and \cite{ORaf} for a review on \cw-algebras in the context
of Hamiltonian reduction.
 We start with a \cw(\cg,\ch) algebra. A
realization of this algebra can be obtained from the Hamiltonian reduction
of the affine Lie algebra
\cg\ w.r.t. the constraints
associated to the \cg-subalgebra \ch. More precisely, starting with the
principal \su\ in \ch,
$\{ E_-,H,E_+\}$, we first
define the gradation of \cg\ w.r.t. the Cartan generator $H$:
\be
\cg=\oplus_{i =-h}^h\ \cg_i =\cg_-\oplus \cg_0\oplus \cg_+
\mb{with} [H, X_i]=i\ X_i \ \ \ \forall X_i\in\cg_i
\ee
Then, we impose the following constraints on a general Lie algebra element \cg:
\be
\left. J(z)\right|_{<0}= E_- \label{primCont}
\ee
i.e. the negative grade part of $J(z)$ is equal to $E_-$.
At the classical level, these constraints generate gauge transformations
\[
J\ \rightarrow\ J^g=gJg^{-1} + k(\prt g)g^{-1} \mb{with} g\in G_+ \mb{and}
\mbox{Lie}(G_+)=\cg_+
\]
and the $W$ generators can be viewed as a basis for the gauge invariant
polynomials. The Poisson
brackets of the original affine Lie algebra \cg\ induce a symplectic
structure on the space of invariant
polynomials, hence providing a realization of the \cw-algebra using Poisson
brackets.

At the quantum level, we have to use ghosts and a BRST operator to take into
account the constraints
(\ref{primCont}). We find the $W$ generators as generators of the zeroth
cohomological space
$H_0(\Omega, s)$.
 $\Omega$ is the enveloping algebra of $\ca=\cg\otimes\Gamma$, where
$\Gamma$ is the algebra
 generated by the ghosts (and anti-ghosts), and $s$ the BRST operator.
$H_0(\Omega, s)$
 possesses  an  algebraic  structure  which  is just the quantum version
of the \cw(\cg,\ch)
 algebra \cite{who}.
 In particular, the classical limit of this
 quantum algebra is the \cw(\cg,\ch) algebra as it has been defined in
terms of Poisson brackets.

 Note that {\it a priori}, the fundamental object in the Hamiltonian
reduction is not the \su\ but
 the gradation  $H$ (since it defines the constraints).  However,
 the  \cw-algebras we get are not classified by these different gradations,
but are in one to one
 correspondence with the  $s\ell(2)$ embeddings in \cg\cite{Oraf2}. Thus,
we get classes of
 gradations  that  lead  to the same \cw-algebra,  each class being
represented by the Cartan
 generator of a \su-subalgebra.
 For a given class, two
 different gradations will differ by a $U(1)$ factor that commute with
the whole \su-subalgebra,
 and which will satisfy a non-degeneration condition\cite{U1}. This $U(1)$
factor (say $Y$) leads
 to  non-trivial  informations about the structure of the \cw-algebra, but
taking as grading operator $H$
 or $H'=H+Y$ provides the same \cw-algebra in two different bases. However,
the gauge group (in
 the classical case) or the BRST operator (in the quantum case) will not be
the same, although the
 final  results are identical. This $U(1)$ factor plays also an important
role for the secondary
 reductions\cite{JOE}.

 \indent

 Now, instead of starting with an affine Lie algebra, and then impose
constraints on its generators,
 one can think of starting directly with a \cw(\cg,\ch) algebra and
impose constraints on some of the
 $W$ generators themselves\cite{Wgd}. This can effectively be done
in certain cases, and the
 resulting  algebra is a ({\it a priori}) new \cw-algebra.
 In fact, starting with a \cw(\cg,\ch) algebra such that
 there  exists  another subalgebra $\ch'$ with
$\ch\subset_{reg}\ch'\subset_{reg}\cg$
 (where the subscript $reg$
 indicates that the embeddings are regular), one can find a set of
constraints such that the Hamiltonian
 reduction of \cw(\cg,\ch) leads to the $\cw(\cg,\ch')$
algebra\footnote{Strictly speaking, this has
 being  proved  only  for  \cg=\sln\ in the classical
case\cite{Wgd}, and most of the cases where
 \cg=\sln,    and some other cases with
 \cg=$so(n)$ or $sp(2n)$, in the quantum case\cite{JOE}. There are however
strong arguments
 (and  examples)  in favor of the generality of this statement.}. Let us
add that there is a natural
 gradation on
 \cw(\cg,\ch) such that the constrained generators are just the $W$
generators of negative grades.

 The secondary reductions lead to chains of
 \cw(\cg,\ch) algebras that reproduce the chains of embeddings of the
\cg-subalgebras. Note that in a
 secondary reduction, we express the resulting \cw(\cg,$\ch'$) algebra
in terms of the polynomials
 in  the generators of the starting \cw(\cg,\ch) algebra: this remark
is fundamental for the linearization of
 \cw-algebras.

 \subsection{Linearization of \cw(\cg,\ch) algebras}

 \indent

Roughly speaking, the linearizations we present are just a special case
of secondary reductions. In
fact, when doing a (primary or secondary) Hamiltonian reduction, we
realize the generators of the
resulting algebra only in terms of the positive grade generators of
the starting algebra (because the
negative grades are all constrained). Then, even if the \cw(\cg,\ch)
algebra is not linear, it may
happen that its positive grade subalgebra $\cw(\cg,\ch)_{\geq0}$ is
linear: it depends both of
\cw(\cg,\ch) and of the gradation we choose (i.e. of the $\cw(\cg,\ch')$
algebra we want to obtain).
If $\cw(\cg,\ch)_{\geq0}$ is linear, the secondary reduction
$\cw(\cg,\ch)\rightarrow\cw(\cg,\ch')$
will provide a linearization of $\cw(\cg,\ch')$. The cases where
this happens have been studied in
\cite{JOE}.

For \cg=\sln, they correspond to the starting algebras \cw(\sln, \su),
called quasi-superconformal
algebras. These algebras have generators only of spins 1, $\frac{3}{2}$,
and one
spin 2. The non-linear terms appears exclusively in the fundamental Poisson
brackets
(or OPEs) between two spin
$\frac{3}{2}$ fields. More precisely, the set of spin $\frac{3}{2}$
fields can be divided into two
subsets $S_\pm$ such that $S_+$ and $S_-$ are Abelian, while the
quadratic terms appears in the Poisson brackets
(or OPEs) of one $S_+$-field with one $S_-$-field. Then, if the (secondary)
constraints on
\cw(\sln, \su) are such that all the $S_-$-fields are constant,
$\cw(\sln, \su)_{\geq0}$ will be
obviously linear. This requirement is satisfied most of the times:

\indent

{\it All the $\cw(\sln,\ \oplus_{i=1}^{m}\ s\ell(p_i))$ algebras are
linearizable if we have
$p_1> p_i+1$ ($i=2,\dots,m)$.}

\indent

We give hereafter concrete formulae for the linearization of
(quantum and classical)
$\cw_n$ algebras.

For \cg=$sp(2n)$, the calculation is the same: we start with the
$\cw(sp(2n), \su)$ algebra to
linearize almost all the $\cw(sp(2n),\ch)$ algebra. More precisely:

\indent

{\it The linearization of the $\cw(sp(2n),\ch)$ is possible when the
subalgebra \ch\ takes the form $\ch=\oplus_i\ s\ell(m_i)\ \oplus_\mu\
sp(2n_\mu)$ with
\begin{itemize}
\item
$ m_1\in 2\itN+1$, $m_1>m_i+2$ $(\forall i\neq1)$,
and $m_1> 2n_\mu+2$ $(\forall \mu)$.
\item
$m_1\in 2\itN$, $m_1\geq m_i+2$ $(\forall i\neq1)$,
and $m_1\geq 2n_\mu+2$ $(\forall \mu)$.
\item
 $n_1> \half(m_i+2)$ $(\forall i)$, and $n_1> n_\mu+1$
 $(\forall \mu\neq1)$.
\end{itemize} }

\indent

For \cg=$so(n)$, the procedure is more restrictive. This is mainly due to
the fact in the $\cw(so(n),
s\ell(2))$ algebra, there is no natural $U(1)$ factor that divides in two
the spin $\frac{3}{2}$
generators, preventing us to find a "natural" linearized subalgebra.

\indent

{\it We can linearize only the $\cw(so(m),so(m'))$ algebras when $m=m'=5$ or
6.}

\indent

In all the linearizations we perform, the secondary reduction one has to
perform is of the type
$\cw(\cg,\ s\ell(2))\  \rightarrow\ \cw(\cg,\ \ch)$, except for the
$\cw(so(5),so(5))$ algebra
which is linearized through
$\cw(so(5),so(3))\ \rightarrow\ \cw(so(5),so(5))$,
and the $s\ell(2)$ algebra is embedded into the
"distinguished" subalgebra of $\ch$ ($s\ell(m_1)$ for $\cg=s\ell(n)$,
and $sp(2n_1)$ or
$s\ell(m_1)$ for $\cg=sp(2n)$).

\subsubsection{Linearization of classical $\cw_n$ algebras\label{truc}}

\indent

We start with the $\cw(\sln,\su)$ algebra as it is obtained from the primary
reduction of \sln. The
fields are gathered in an $n\times n$ matrix:
\be
\left(\begin{array}{cc|cccc}
U & T & G_1^+ & G^+_2 & \cdots & G^+_{n-2} \\
1 & U & 0 & 0 & \cdots & 0 \\
\hline
0 & G_1^- & & & &\\
0 & G_2^- & & & &\\
\vdots & \vdots & & {\bf M} &- \frac{2}{n-2}U\cdot\1_{n-2} & \\
0 & G_{n-2}^- & & & &
\end{array}\right)
\mb{with}
\left\{ \begin{array}{ll}
T & \mbox{the spin 2 generator}Ê\\
G_i^\pm & \mbox{the spin $\frac{3}{2}$ generators}Ê\\
U & \mbox{$U(1)$ generator}Ê\\
{\bf M} & \mbox{a $s\ell(n-2)$ matrix}
\end{array}\right. \label{quadra}
\ee
The subsets $S_\pm$ already mentioned are formed with the $G_i^\pm$
generators and correspond
to the eigenvalues of these fields under the action of $U$. From the
form of the above matrix, it is
easy to see that the secondary constraints leading to the $\cw_n$
algebra are
\be
G_i^-=\delta_{i,1} \mb{and} M_{-\alpha}=\chi_\alpha\ \
(\alpha\mbox{ positive root})
\label{2cont}
\ee
where $\chi_\alpha$ are the constraints of the Abelian Toda model
built on $s\ell(n-2)$ (i.e.
$\chi_{\alpha}=1$ if $\alpha$ simple root of $s\ell(n-2)$ and 0 otherwise).
These constraints
generate gauge transformations on the \cw(\sln,\su)-fields.
In \cite{Wgd}, it has been shown that a correct gauge fixing for these
secondary reductions is:
\be
(M_{\geq0})^g=0 \mb{and} U^g=0 \label{gfix}
\ee
A priori, to get the realization of the resulting \cw-algebra, one should
first compute all the Poisson brackets of the
\cw(\sln, \su) algebra and then make the gauge transformations.
Fortunately, for our purpose, there
is a simpler way to get the realization.

The idea is to consider the general group transformations associated
to \cn, the algebra spanned by
the positive root generators:
\be
J\ \rightarrow\ J^g=gJg^{-1} + k(\prt g)g^{-1} \mb{with} g= \exp
\left(\begin{array}{cccc} 0 &* & \dots & * \\ \vdots &\ddots & \ddots &
\vdots \\
\vdots & & \ddots & * \\ 0 &\dots & \dots & 0 \end{array}\right)
\label{jens}
\ee
For a general element $g$, (\ref{jens}) do not respect the form
(\ref{quadra}) with constraints
(\ref{2cont}) assumed. However, demanding the transformations (\ref{jens})
to send the matrix
(\ref{quadra}) with constraints (\ref{2cont}) to a matrix of the form:
\be
\left(\begin{array}{ccccc}
0 &W_1 & \dots & W_{n-2} & W_{n-1} \\
1 &0 & \dots & 0 & 0 \\
0 & 1 &  & 0 & 0 \\
\vdots & & \ddots & & \vdots \\
0 &0 & \dots & 1 & 0
\end{array}\right)
\ee
completely fixes the parameters of $g$ and give the expression of the $W_i$
in terms of the elements
$T(w)$, $G_i^+(w)$, and $M_{\geq0}(w)$, $U(w)$.

Thus it is a matter of straightforward calculations to get an expression for
the parameters
in $g$ such that (\ref{gfix}) is satisfied. Once this is done, the
expressions of $[T(w)]^g$
and $[G_i^+(w)]^g$ will provide the linearization of the $\cw_n$ algebra
in terms of the fields
$T(w)$, $G_i^+(w)$, and $M_{\geq0}(w)$, $U(w)$. Note that we only need to
know the Poisson brackets (or
OPEs) in the linearized algebra, not in the full \cw(\sln, \su) algebra.
The calculation of these OPEs
has been replaced by extra equations coming from the general
transformations (\ref{jens}).

To be complete, we give hereafter the OPEs of $\cw(\sln, \su)_{\geq0}$:
\be
\begin{array}{llll}
T(z)T(w)= & \frac{c/2}{(z-w)^4} +2\frac{T(w)}{(z-w)^2}
+\frac{\prt T(w)}{(z-w)}\ &
T(z)U(w)=&  \frac{U(w)}{(z-w)^2} +\frac{\prt U(w)}{(z-w)} \\ \\
T(z)M_a(w)= & \frac{M_a(w)}{(z-w)^2} +\frac{\prt M_a(w)}{(z-w)} &
T(z)G_i^+(w)= & \frac{3}{2}\frac{G_i^+(w)}{(z-w)^2}
+\frac{\prt G_i^+(w)}{(z-w)} \\ \\
M_a(z)M_b(w)= & \frac{k g^{ab}}{(z-w)^2} +f_{ab}^c\frac{M_c(w)}{(z-w)} &
U(z)U(w)=&  \frac{k'}{(z-w)^2}\\ \\
M_a(z)G_i^+(w)= & F_{ai}^j\frac{G_j^+(w)}{(z-w)} &
U(z)G_i^+(w)=& \frac{G_j^+(w)}{(z-w)}\\ \\
G_i^+(z)G_j^+(w)= & \mbox{regular} & U(z)M^a(w)=& \mbox{regular}
\end{array}
\ee

\subsubsection{Linearization of quantum $\cw_n$ algebras\label{linqa}}

\indent

For the quantum version of the above linearization, we need to calculate
the cohomology space of a
given BRST operator. The calculation is very similar to the primary
reduction case \cite{dBT}. We
first have to introduce a pair of ghosts for each constraints (\ref{2cont}):
$(c_i,b^i)$ $i=1,\dots,
n-2$ and $(c_\alpha,b^\alpha)$, $\alpha$ positive root. Then, the BRST
operator acts as
\[
\begin{array}{l}
s(\vph(w))= \oint_w dz\ j(z)\vph(w) \mb{with} \\
j(z)=(M_\alpha(z)-\chi_\alpha)c^\alpha(z)+
(G^-_i(z)-\delta_{i,1})c^i(z)+ \half\ {f_{\alpha\beta}}^\mu
(b_\mu c^\beta c^\alpha )_0(z)
+ {f_{\A i}}^j (b_j c^i c^\A)_0(z)
\end{array}
\]
where $(\ )_0$ denotes the normal ordering $(AB)_0(w)=\oint_w\ \frac{dz}{z-w}
A(z)B(w)$.
This operator can be divided into two parts $s=s_0+s_1$, and a tic-tac-toe
procedure leads to the
$W$ generators. More precisely, we first divide
the set $\Omega$  in two
subsets $\wh{\Omega}$ and $\cb$ such that $\Omega=\wh{\Omega}\otimes\cb$,
$s(\cb)\subset\cb$, $H(\cb,s)=\C$, and $\wh{\Omega}$ satisfying
$s(\wh{\Omega})\subset\wh{\Omega}$, which leads to
$H_*(\Omega, s)=H_*(\wh{\Omega}, s)$.

In practice, \cb\ is built on the ghosts $b^i$, $b^\alpha$ and $s(b^i)$,
$s(b^\alpha)$ , while
$\wh{\Omega}$ is generated by a suitable redefinition of the fields
$J^{\bar{\A}}$, $G^i$
and $T$ (see below).

Then, defining $j_0(z)=-\chi_\alpha c^\alpha(z)-\delta_{i,1}c^i(z)$
and $j_1(z)= j(z)-j_0(z)$, one starts by proving that {\it as
vector spaces} there is an isomorphism between $H_0(\wh{\Omega}, s)$ and
$H_0(\wh{\Omega},s_0)$. But,  $H_0(\wh{\Omega},s_0)=\mbox{ker }
s_0|_{\wh{\Omega}}$ which
we denote by $\wh{\Omega}_0$.
Note that the
fields of $\wh{\Omega}_0$ play the role of the highest weights in the primary
reductions.
The generators of
the \cw-algebra are obtained through the recursive relations
$s_1(W_i)=s_0(W_{i+1})$,
$W_0\in\wh{\Omega}_0$,
and $W=\sum_i\ (-)^i W_i$.
There is a bi-gradation (built on the initial gradation and the ghost number)
that
ensures that the sum is
finite. $W_0$ has a bi-grade $(p,-p)$
and at each step $W_i$ has a definite bi-grading $(p-i,i-p)$, with
$i=0,\dots, p$.
Thus, the only technical difficulty is
to find a set of generators $W_0$ for $\wh{\Omega}$.
The following procedure provides a simple and explicit realization for
$\wh{\Omega}$ generators
in a finite number of steps:

We start with  the $J^{\bar\A}$, $G^i$ and $T$ generators.
 The generators $J^{\bar\A}$
 are just residual Kac-Moody currents, so that the corresponding
"hatted" generators are known:
\be
\wh{J}^{\bar\A}=J^{\bar\A} + {f^{{\bar\A}d}}_e\ (b^e c_d)_0
\ee

Now, we focus on the $G^i$'s.
$s(G^i)$ can be decomposed as
$s(G^i)= R^i_{\A\B}(\wh{J},c) J^\A J^\B+P^i_\A(\wh{J},c)J^\A
+S^i_\A(\wh{J},c)\prt J^\A
+Q^i(\wh{J},c)$.  We first define
$G^i_1= \half R^i_{\A\B}(\wh{J},c)\, (J^\A b^\B+b^\A J^\B)
+P^i_\A(\wh{J},c)b^\A
+S^i_\A(\wh{J},c)\prt b^\A$  and compute $s(G^i-G^i_1)$:
as a differential polynomial in the  constrained currents (say $W^\A$) and
$b$'s, it contains terms of the form $W^\A b^\B$, $W^\A$, and $\prt W^\A$ .
Then, we introduce a $G^i_2$,
which is just given by $s(G^i-G^i_1)$ with the replacement rules
$W^\A b^\B\ \rightarrow\ \half b^\A b^\B$, $W^\A\ \rightarrow\ b^\A$,
$\prt W^\A\ \rightarrow\ \prt b^\A$ and any term which does not contain any
$W^\A$ is replaced by
0 (see definition of $G^i_1$ starting from $s(G^i)$). The final expression
for $\wh{G^i}$ is
$\wh{G^i}=G^i-G^i_1-G^i_2$.

Finally, $T$ can be computed explicitly for any $n$; it reads:
\be
\wh{T}= T-\frac{3}{2} (b_i\prt c^i)_0
+ \half (\prt b_i\, c^i+ b_\A\prt c^\A)_0.
\ee

The above procedure can be generalized to any linearizable \cw-algebra: see
the appendix for the
scheme of linearization and the proofs.

We have checked the expressions for $n=2,3,4$ and 5. Note that for $n$
bigger or equal to 5,
there is a technical fact to take into account: if one computes the OPEs
of the above generators using
Kac-Moody OPEs, one realizes that $\wh{\Omega}$ is {\em \underline{not}}
an algebra. New
generators, built only on ghosts, but with total ghost-number 0 appear at
the right hand side of the
OPEs. To
make $\wh{\Omega}$ an algebra, one has to {\em define} new OPEs through
the formula:
\be
A\diamond B= \pi(A\circ B)\ \ \forall\ A,B\in  \wh{\Omega}
\mbox{ with }\pi \mbox{ the
projection onto } \wh{\Omega}
\ee
where as a notation we have used $(A\circ B)$ to denote the OPE of
$A$ with $B$.

Note that we have not rigorously proven that ``$\diamond$'' does in fact give
an associative algebra in the general case, but
we give a number of arguments for the validity of the procedure
in the appendix.

Let us also remark that an alternative (and less systematic) approach for
the linearization of \cw-algebras exists. In this approach, we start from
the classical linearization (obtained as above). We adjust the various
coefficients (in the expressions for the \cw-generators and in the
expressions for the OPEs of the linearizing algebra) in such a way that the
quantum OPEs of the \cw-generators gives a closed algebra. In that case
clearly no extra ghost-terms appear, and it is clear that we do not need to
modify the OPEs: everything will work as in the classical case.
It seems evident that the linearizing algebra obtained in this way is
identical to the algebra defined by the ``$\diamond$'' composition.
\sect{Linearization of \cw-superalgebras}

\indent

The case of super \cw-algebras is very similar to what we have
described so far for bosonic
\cw-algebras. One has first to study secondary reductions for
super \cw-algebras, and
then to separate the reductions that provide
linearizations. We first recall some general features on the Hamiltonian
reductions of superalgebras.

\subsection{Generalities on Hamiltonian reduction of superalgebras}

\indent

 As for algebras, to perform a Hamiltonian reduction on a superalgebra \sg,
we need to consider a
 gradation $H$ that will ensure the nil-potency of our set of (first class)
constraints. The different
 \cw-superalgebras one can get are not classified by the different gradations,
but more precisely by the
 \su\ embeddings: there is still a freedom in a shift by a $U(1)$ factor.
 The classification of the \su-embeddings in
 \sg\ is of same type as for Lie algebra: each \su\ algebra can be seen as
the principal embedding in a
 regular subalgebra of \sg, up to some exceptions for the $osp(m|n)$
superalgebras. Note that the
 subalgebra is embedded in the bosonic part of \sg:
 although they have strong effects in the Hamiltonian reduction, the
fermions do not play any role in
 this  classification. However, a super-symmetric treatment of Hamiltonian
reduction can also be
 done \cite{DRS,FRS,JOE}: in that case, the super-symmetric \cw-algebras are
classified by
 $osp(1|2)$ embeddings, and the fermions do enter in the game.

 Once the gradation is determined, the reduction follows the same lines as
the bosonic case. The only
 difference relies on the grades $\pm\half$ (when they exist). Indeed, the
$\cg_\half$ part leads to
 second class constraints: to cure it in Lie algebra, one has to use a
"halving" procedure or to shift
 the grading operator with a $U(1)$ factor. For superalgebras, these
techniques are not always
 possible, and in general one adds free (fermionic) spin $\half$ fields
(see examples).

 At the classical level, one has just to constrain the $\cg_{<-\half}$ part
of the current $J$ as usual,
 constrain the $\cg_{-\half}$ part to be the free fields, and then fix the
gauge symmetry in the
 highest weights gauge for instance.

 At the quantum level, the free fields induce a new term in the
 decomposition of $s$: $s=s_{(0,1)}+s_{(\half,\half)}+s_{(1,0)}$ where the
indices refer to the
 bi-grading ($s_{(0,1)}$ is the $s_0$ part of section \ref{linqa}, while
$s_{(1,0)}$ is $s_1$).
 Although  more complicated, the calculation can still be done\cite{SeThTr},
and a tic-tac-toe
 construction used to get the corresponding generators.

 \indent

 For the secondary reductions of \cw-superalgebras, we have to find a gauge
fixing that ensures the
 embeddings of the sets of constraints. As far as $s\ell(m|n)$ superalgebras
are
 concerned, the calculation is very similar to the case of $s\ell(m+n)$
algebras. Thus, we can define a
 generalized horizontal gauge for the classical case, or tune a $U(1)$ factor
for the quantum case.

 More precisely, at the classical level, if one wants to reduce $s\ell(m|n)$
with respect to, for
 instance,
 $[s\ell(m_1)\oplus s\ell(m_2)]\oplus s\ell(n_1)$ (with $s\ell(m_i)$ in
$s\ell(m)$ and $s\ell(n_1)$ in
 $s\ell(n)$),
 the  gauge  fixing  will  be  of  the  form
 \be
 \left(\begin{array}{c||c}
 \begin{array}{ccc|cccc|cc}
  * & * & * &  \phantom{0} & * & * & * & * &* \\
 1 & (m_1) &  \phantom{0} &  \phantom{0} &  \phantom{0} &
  \phantom{0} &  \phantom{0} &  \phantom{0} &  \phantom{0} \\
  \phantom{0} & 1 &  \phantom{0} &  \phantom{0} &  \phantom{0} &
   \phantom{0} &  \phantom{0} &  \phantom{0} &  \phantom{0} \\
 \hline
  \phantom{0} &  \phantom{0} & * & * & * & * & * & * &* \\
  \phantom{0} &  \phantom{0} & * & 1 & (m_2) &  \phantom{0} &  \phantom{0} &
  \phantom{0} &  \phantom{0} \\
  \phantom{0} &  \phantom{0} & * &  \phantom{0} & 1 &  \phantom{0} &
\phantom{0} &
   \phantom{0} &  \phantom{0}  \\
  \phantom{0} &  \phantom{0} &  \phantom{0} &  \phantom{0} &  \phantom{0} &
1 &
   \phantom{0} &  \phantom{0} &  \phantom{0}  \\
 \hline
  \phantom{0} &  \phantom{0} & * &  \phantom{0} &  \phantom{0} &  \phantom{0}
& * & * &* \\
  \phantom{0} &  \phantom{0} & * &  \phantom{0} &  \phantom{0} &  \phantom{0}
& * & * & * \\
 \end{array}\
 &\
\begin{array}{ccc|cc}
* & \hspace{.5em}*\hspace{.5em}  & * & * & * \\
 \phantom{0} &  \phantom{0} &  \phantom{0} &  \phantom{0} &  \phantom{0} \\
 \phantom{0} &  \phantom{0} &  \phantom{0} &  \phantom{0} &  \phantom{0} \\
\hline
* & * &* & * & * \\
\phantom{0} & \phantom{0} & \phantom{0} & \phantom{0} & \phantom{0} \\
\phantom{0} & \phantom{0} & \phantom{0} & \phantom{0} & \phantom{0} \\
\phantom{0} & \phantom{0} & \phantom{0} & \phantom{0} & \phantom{0} \\
\hline
 \phantom{0} & * & * & * & * \\
 {\phantom{0}} & \phantom{0} & \phantom{0} & * & * \\
\end{array}
\\ \\[-1.4mm]
\hline \hline
\\[-1.4mm]
\begin{array}{ccc|cccc|cc}
 \phantom{0} & \hspace{.7em}\phantom{0}\hspace{.7em}  & * & \phantom{0}
 & \hspace{.7em}\phantom{0}\hspace{.7em} & \phantom{0} & * & \phantom{0} &
* \\
 \phantom{0} & \phantom{0} & * & \phantom{0} & \phantom{0} & \phantom{0} & * &
 \phantom{0} & * \\
 \phantom{0} & \phantom{0} & * & \phantom{0} & \phantom{0} & \phantom{0} & * &
 \phantom{0} & \phantom{0} \\
 \hline
 \phantom{0} & \phantom{0} & * & \phantom{0} & \phantom{0} & \phantom{0} & *
& * &* \\
 \phantom{0} & \phantom{0} & * & \phantom{0} & \phantom{0} & \phantom{0} & *
& * & * \\
 \end{array}\
 &\
 \begin{array}{ccc|cc}
 * & * & * & * & * \\
 1 & (n_1) & \phantom{0} & \phantom{0} & \phantom{0}  \\
 \phantom{0} & 1 & \phantom{0} & \phantom{0} & \phantom{0} \\
 \hline
 \phantom{0} & \phantom{0} & * & * & * \\
 \phantom{0} & \phantom{0} & * & * & *
 \end{array}
 \end{array}\right) \label{ghg}
 \ee

 \noindent
 where the double lines delimit fermions and bosons, while the single lines
indicate the positions of
 the $s\ell(m_i)$ and $s\ell(n_1)$ subalgebras. The entries where a $W$
generator appears are
 indicated by a star ($*$).
 For more details about the generalized horizontal
 gauge, see \cite{Wgd}. This proves that the secondary reductions for the
\cw\ superalgebras
 based on $s\ell(m|n)$  are always possible (at the classical level). For
$osp(m|2n)$ superalgebras,
 such a horizontal gauge does not exist, but numerous examples (at the
classical level) indicates that
 the secondary reductions are possible at least when the subalgebras that
define the different
 $s\ell(2)$ embeddings are taken regular (which is not always possible,
contrarily to the $s\ell(m|n)$
 case). Thus, we can say:

 \indent

 {\em At the classical level, we can perform the secondary reductions:
 \begin{description}
 \item[ ]  $\cw(s\ell(m|n),\ch)\ \rightarrow\  \cw(s\ell(m|n),\ch')$ as soon
as $\ch\subset\ch'$.
 \item[ ]
  For $osp(m|2n)$ superalgebras, there is no doubt
 that $\cw(osp(m|2n),\ch)\ \rightarrow\ \cw(osp(m|2n),\ch')$ is possible as
soon as
 $\ch\subset_{reg}\ch'$
 \end{description} }

 \indent

  At the quantum level, we have to look at the sets of first class
constraints of the different reductions,
  and  see whether they can be embedded one into each other. The
calculation is rather cumbersome:
  using $U(1)$ factors if necessary, we have to check case by case if
the sub-superalgebra $\sg_+$
  is embedded in $\sg'_+$. At the end,
  we are led to the result:
\newpage
  \indent

 {\em At the quantum level, we can perform the secondary reductions of the
type:
 \begin{description}
 \item[ ]
 $\cw(s\ell(m|n),\ch)\ \rightarrow\  \cw(s\ell(m|n),\ch')$
 with $\ch=\oplus_i s\ell(m_i)$ and
 $\ch'=\oplus_\mu s\ell(m'_\mu)$  as soon as
 $|m_i-m_j|>|m'_\mu-m'_\nu|$ $\forall i,j,\mu,\nu$ with $i\neq j$ and
$\mu\neq\nu$.
 \item[ ]
 $\cw(osp(m|2n),\ch)\ \rightarrow\ \cw(osp(m|2n),\ch')$ is possible when
 \begin{itemize}
 \item $\ch=s\ell(2)\subset sp(2N)$ and
 $\ch'=\ch'_1\oplus\ch'_2$ where
$\ch'_1=[\oplus_i so(m_i)\oplus_\mu s\ell(m_\mu) ]\subset
 so(M)$,    $\ch'_2=[\oplus_j sp(2n_j)\oplus_\nu s\ell(p_\nu) ]\subset
sp(2N)$ obey to
 one  of  the three conditions
       \begin{enumerate}
       \item
       $n_1> n_j+2$ $(\forall j\neq1)$,  $n_1>\half(m_\mu+2)$ $(\forall \mu)$,
       $n_1\geq \left[\frac{m_i-1}{2}\right]+2$ $(\forall i)$
       and $n_1>\half(p_\nu+2)$ $(\forall \nu)$.
       \item
      $p_1\in 2\itN$, $p_1\geq p_\nu+2$ $(\forall \nu\neq1)$,
$p_1\geq m_\mu+2$ $(\forall \mu)$,
       $p_1\geq 2\left[\frac{m_i-1}{2}\right]+3$ $(\forall i)$
      and $p_1\geq 2n_j+2$ $(\forall j)$.
      \item
        $p_1\in 2\itN+1$, $p_1> p_\nu+2$ $(\forall \nu\neq1)$,
$p_1> m_\mu+2$ $(\forall \mu)$,
       $p_1> 2\left[\frac{m_i-1}{2}\right]+3$ $(\forall i)$
      and $p_1> 2n_j+2$ $(\forall j)$.
       \end{enumerate}
 \item $\ch=s\ell(2)\subset sp(2N)$ and
 $\ch'=\ch'_1\oplus sp(4)$ where
 $\ch'_1=[m_1\ s\ell(3)\oplus m_2\ so(3) ]\subset so(M)$
 \item $\ch=s\ell(2)\subset sp(2N)$ and
 $\ch'=\ch'_1\oplus\ch'_2$ where $\ch'_1=m\ s\ell(2)\subset so(M)$, and
 $\ch'_2=(sp(4)\oplus_\nu s\ell(p_\nu) )\subset sp(2N)$
\end{itemize}
 \item[ ]
 We conjecture that the general secondary reduction
 $\cw(\sg,\ch)\ \rightarrow\ \cw(\sg,\ch')$ will be possible as soon as
$\ch\subset_{reg}\ch'$.
 \end{description} }

\indent

Note that what we have presented is a classification of algebras in which
the secondary reductions can be carried out using our procedure of modifying
the
set of first class constraints by
adding $U(1)$ currents to the grading. It is not intended as an exhaustive
classification of algebras
where a secondary hamiltonian reduction is possible.
On the contrary, as mentioned above, we conjecture that
the secondary reductions are possible as soon as the embeddings are regular.
However, to show this
general property, other methods will have to be used, while the "$U(1)$
factor technique" not only
ensures the secondary reduction but also provide an explicit realization for
it.

As we are interested here only in the linearization of \cw-superalgebras, we
will not go
further into details. The proofs of the above properties are very similar to
the bosonic case, and we
refer to \cite{JOE} for the interested reader.  We rather focus on the
linearization.

\subsection{Linearizations}

\indent

The linearization of \cw-superalgebras is very similar to the linearization
of \cw-algebras.
For a given
\cw(\sg, \ch) superalgebra, and among all the possible secondary reductions,
we look for
$\cw(\sg, \ch_0)$
superalgebra such that the reduction $\cw(\sg, \ch_0)
\rightarrow\cw(\sg, \ch)$ is possible and
$\cw(\sg, \ch_0)_{\geq0}$ is a linear superalgebra.

\subsubsection{$s\ell(m|n)$ superalgebras}

\indent

As already mentioned, the reduction of $s\ell(m|n)$ superalgebras is
very similar to the case of
$s\ell(m+n)$ algebras. Indeed, for the linearization, the results are
still very similar:

\indent

{\em The linearization of $\cw[s\ell(n|m), \ch]$ superalgebras can be done
by the secondary
reductions $\cw(s\ell(n|m), s\ell(2))\rightarrow\cw(s\ell(n|m), \ch)$
as soon as \ch\ satisfies:
$\ch=\oplus_i\ s\ell(p_i)$ with $p_1>p_i+1$ ($i\neq1$). In that case,
the $\ch_0=s\ell(2)$ algebra
will be embed in $s\ell(p_1)$ for the secondary reduction.

In particular, the superalgebras $\cw[s\ell(n|m), s\ell(n)\oplus s\ell(m)]$
are linearizable as soon as
$m \neq n,\ n\pm1$.}

\indent

Let us first describe the $\cw(s\ell(n|m), s\ell(2))$ superalgebra: it
contains one spin 2 generator $T$
(stress energy tensor), $2m$ fermionic spin $\frac{3}{2}$ fields $G_i^\pm$
(super-symmetry charges), a
$s\ell(n-2)\oplus s\ell(m)\oplus g\ell(1)$ Kac-Moody subalgebra (spin 1
generators $M_a$),
as well as
$2n(m-2)$ spin 1 fermionic fields $Q^\pm_{ij}$ and $2m$ spin $\frac{3}{2}$
bosonic fields
$D_i^\pm$.

The fields $G_i^\pm$ and the fields $D_i^\pm$
form a $m\oplus\bar{m}$ representation of the $s\ell(m)$
subalgebra, they are trivial under $s\ell(n-2)$ transformations, and the
index $\pm$ denotes the
$g\ell(1)$ charge. The fields $Q^\pm_{ij}$ form two $(n-2,m)$
representations of
$s\ell(n-2)\oplus s\ell(m)$.

\indent

At the classical level, for the linearization, we start with the
$\cw(s\ell(n|m), s\ell(2))$ superalgebra
in the generalized horizontal gauge (as defined above, see example
(\ref{ghg})), and we perform a
gauge transformation that will lead to the $\cw[s\ell(n|m), \ch]$
superalgebra in the generalized
horizontal gauge. The expression of the transformed fields as functions of
the fields of
$\cw[s\ell(n|m), \ch]_{\geq0}$ provides the linearization. Note that the
techniques described in section
\ref{truc} still works.

At the quantum level, we introduce the index $\alpha$ and $\bal$ for the
respectively negatively and
non-negatively  graded Kac-Moody generators. Then, the BRST current is
\be
\begin{array}{ll}
j(z)= & (M_\alpha(z)-\chi_\alpha)c^\alpha(z)+ (G^-_i(z)-\chi_i)\gamma^i(z)+
(Q^-_{ij}(z)-\chi_{ij})\gamma^{ij}(z)+ (D^-_i(z)-\chi_i)c^i(z)+\\
& + \half\ \eps_b{f_{ab}}^c (B^cC_bC_a)_0(z)
\end{array}
\ee
where we have introduced fermionic ghosts $(b^\alpha, c_\alpha)$ and
$(b^i,c_i)$, and bosonic
ghosts $(\beta^{ij}, \gamma_{ij})$. We have denoted generically by
$B$ and $C$ the ghosts and
by $a,b,c,\dots$ the indices; ${f_{ab}}^c$ are the structure constants of
the (linear) algebra
$\cw_{\geq0}$, and $\eps_b=1$ (resp.
$\eps_b=-1$) if $C_b$ is a bosonic (resp. fermionic) ghost.
The cohomology of $s$ will provide the linearization.
We present hereafter an example for the calculation of this cohomology
(hence for the linearization).

\subsubsection{$osp(M|2N)$ superalgebras}

\indent

The calculation for $osp(M|2N)$ superalgebras resembles the one for
$s\ell(m|n)$. Using the results
for $so(M)$ and $sp(2N)$ algebras, one sees that we will mostly use the
$sp(2N)$ part to linearize,
except for few particular values of $M$ in $so(M)$. More precisely:

\indent

{\em The linearization of  $\cw[osp(2M+1|2N), \ch]$ superalgebras is possible
through the secondary
reduction $\cw[osp(2M+1|2N), \ch_0]\rightarrow\cw[osp(2M+1|2N), \ch]$ when
$\ch_0=s\ell(2)\subset sp(2N)$ and
$\ch=\oplus_p\ so(m_p)\ \oplus_i\ s\ell(m_i)\ \oplus_\mu\ sp(2n_\mu)$ with
\begin{itemize}
\item Either
$m_1\in 2\itN$, $m_1\geq m_i+1$ $(\forall i\neq1)$,  $m_1\geq 2m_p+1$
$(\forall p)$,
and $m_1\geq 2n_\mu+1$ $(\forall \mu)$.
\item Or
 $n_1\geq \half(m_i+1)$ $(\forall i)$, $n_1\geq m_p+1$ $(\forall p)$, and
$n_1\geq 2n_\mu+1$
 $(\forall \mu\neq1)$.
\end{itemize}
In particular, the linearization of $\cw[osp(2m+1|2n), so(2m+1)\oplus
sp(2n)]$ superalgebras is
possible when $m<n-1$.}

\indent

As for $s\ell(m|n)$ superalgebras, the linearization is done using a
quadratic superconformal algebra.
The calculations are of same type: for the classical level, we have to
compute the gauge
transformations that leads to the (secondary) \cw-superalgebra, and at the
quantum level, it is once
more the cohomology of the BRST operator that will give rise to the
linearization. The new feature is
the emergence of auxiliary fields: see example below.

\sect{Examples}
\subsection{Case of $\cw(s\ell(3|1), s\ell(3))$}
\subsubsection{Classical Linearization}

\indent

In order to demonstrate the linearization procedure, we will consider the
linearization of the algebra $\cw(s\ell(3|1),s\ell(3))$ in some detail.
This superalgebra
is comparatively simple, but it still shows most of the characteristics
of our linearization procedure. Note that this example has already been
done in \cite{BKS}; we repeat it here only to demonstrate our method.

We start with the superalgebra $s\ell(3|1)$, parameterized by
$$
J(z) = \left ( \begin{array}{rrr|r}
       H_1(z)+Y(z) &              J_{1}(z)&            J_{2}(z)& j_{1}(z)\\
           J_{4}(z)&H_{2}(z)-H_{1}(z)+Y(z)&            J_{3}(z)& j_{2}(z)\\
           J_{5}(z)&              J_{6}(z)&      -H_{2}(z)+Y(z)& j_{3}(z)\\
\hline
\bar{\jmath}_{1}(z)&   \bar{\jmath}_{2}(z)& \bar{\jmath}_{3}(z)& 3Y(z)
\end{array}
\right )
$$
The constraints and highest weight gauge resulting in the algebra
$\cw(s\ell(3|1),s\ell(2))$ are
\bea
J_c(z) & = & \left ( \begin{array}{rrr|r}
       H_1(z)+Y(z) &              J_{1}(z)&            J_{2}(z)& j_{1}(z)\\
                  1&H_{2}(z)-H_{1}(z)+Y(z)&            J_{3}(z)& j_{2}(z)\\
                  0&              J_{6}(z)&      -H_{2}(z)+Y(z)& j_{3}(z)\\
\hline
                  0&   \bar{\jmath}_{2}(z)& \bar{\jmath}_{3}(z)& 3Y(z)
\end{array}
\right ) \nn
\vspace{5mm} \nn
 J_{hw}(z) & = & \left ( \begin{array}{rrr|r}
     U(z)+Y(z)&    \tilde{T}(z)&     W^+(z)& G^+(z)\\
             1&       U(z)+Y(z)&          0&      0\\
             0&          W^-(z)&-2U(z)+Y(z)& B^-(z)\\ \hline
             0&          G^-(z)&     B^+(z)&  3Y(z)
\end{array}\right ).
\ena
The secondary first class constraints that leads to the algebra
$\cw(s\ell(3|1),s\ell(3))$ are $\bar{W}(z) =1$ and $\bar{G}(z)=0$, and these
first
class constraints induce a gauge-invariance which can be used to choose the
gauge $U(z)=B(z)=0$, so the result is
\be
\label{Jw}
J_{w}(z) = \left ( \begin{array}{rrr|r}
  \cy(z)&  \tilde{\ct}(z)&        W(z)&         G(z)\\
             1&    \cy(z)&           0&            0\\
             0&               1&\cy(z)&            0\\ \hline
             0&               0&        B(z)& 3\cy(z)
\end{array}\right ).
\ee

$U$ and $Y$ are two $U(1)$ Kac-Moody currents (spin 1 primary fields), $W^\pm$
are primary spin $\thf$ bosonic fields with $(U,Y)$ charges
$(\pm\hf,0)$, $G^\pm$ are primary spin $\thf$ fermionic fields with charges
$(\pm\frac{1}{6},\pm \th)$ and $B^\pm$ are fermionic Kac-Moody currents
with charges $(\pm \th, \mp \th)$.

Again, doing the finite gauge transformation
$J \rightarrow J^g = gJg^{-1} + k (\pa g)g^{-1}$ with $g=\exp(\Lambda)$ and
$$
\Lambda = \left (  \begin{array}{rrr|r}
             0 &   \eps_1&  \eps_2 &   \G_1  \\
             0 &        0&       0 &      0  \\
             0 &   \eps_3&       0 &      0  \\ \hline
             0 &     \G_2&       0 &      0  \end{array} \right )
$$
and requiring the result to respect the secondary constraints and take the
form (\ref{Jw}), we find
conditions for all the gauge parameters, and the result $J^g$ gives the
classical linearization of $\cw(s\ell(3|1),s\ell(3))$:
\bea
\label{lin-cl}
\ct & = & \frac{1}{k} \left ( T + 3Y^2 - (3k-1) \pa U + \pa Y \right ) \nn
W   & = & W^+ - k\, B^- \pa B^+ - 2 k\, \pa B^-B^+ +
   2 k T U + (6k+2) U \pa U + 2U \pa Y+ \nn
& &  + 4B^- B^+ U - 2B^- B^+ Y -
   8U^3 + 6U Y^2 - 2k^2\, \pa^2 U \nn
G & = & G^+ - k\, B^- T - B^- \pa U - k\, B^- \pa U -
   B^- \pa Y + 2k\, B^- \pa Y -  2k\, U\pa B^- +  \nn
& &     + 4k\, Y\pa B^- + 4B^- U^2 -
   4B^- U Y + B^- Y^2 + k^2\, \pa^2 B^- \nn
B & = & B^+ \nn
\cy & = & Y
\ena
Note that $T$ and $\ct$ are the normalized
energy-momentum tensors, corrected with quadratic terms in the Kac-Moody
currents, such that all fields are primary.

\subsubsection{Quantum Linearization}

In order to perform the quantum linearization, we need to know the operator
product expansions of the algebra $\cw(s\ell(3|1),s\ell(2))$. As noted above,
this
algebra consists of the energy-momentum tensor $T$ with central charge
$-\frac{(2k+1)(3k+4)}{k+2}$,
two $U(1)$ Kac-Moody currents $U$ and $Y$, two spin $\thf$ bosonic fields
$W^\pm$ with $(U,Y)$ charges
$(\pm\hf,0)$, two fermionic spin $\thf$ fields $G^\pm$ with charges
$(\pm\frac{1}{6},\pm \th)$, and fermionic Kac-Moody currents $B^\pm$
with charges $(\pm \th, \mp \th)$.
The non-trivial operator product expansions not already implicitly given are:
\newpage
\bea
\ope{U}{U} & = & \ord{2}{(3k+4)/18} \\
\ope{U}{Y} & = & \ord{2}{-1/18}  \nn
\ope{Y}{Y} & = & \ord{2}{-(3k+2)/18} \nn
\ope{B^+}{B^-} & = & \ord{2}{-(k+1)}+ \ordo{-2Y-2U} \nn
\ope{B^\pm}{G^\pm} & = & \ordo{W^\pm}   \nn
\ope{B^\pm}{W^\mp} & = & \ordo{\pm G^\mp} \nn
\ope{W^\pm}{G^\mp} & = & \ord{2}{2(k+1)B^\pm} +
\ordo{ \pm 4 (B^\pm U)_0 \mp 2 (B^\pm Y)_0 + (k+2)\pa B^\pm} \nn
\ope{W^+}{W^-} & = & \ord{3}{-2(1+k)^2} + \ord{2}{2Y-(6k+4)U} +\nn
& & +\ordo{(k+2)T + 2 (B^-B^+)_0 - 12 (UU)_0 + 3 (YY)_0 - 3k \pa U+ 3 \pa Y}
\nn
\ope{G^+}{G^-} & = & \ord{3}{-2(1+k)^2} + \ord{2}{-2kU+(4k+6)Y} +\nn
& & +\ordo{(k+2)T - 4 (UU)_0 + 4(UY)_0 - (YY)_0 - k\pa U + (2k+3) \pa Y}
\nonumber
\ena

We now introduce a fermionic ghost pair $(b,c)$ corresponding to the
secondary first class constraint $W^- = 1$, and a bosonic ghost pair $(\B,\G)$
corresponding to the constraint $G^-=0$, and we define the BRST current
$$
j = (W^--1)c + G^-\gamma
$$
As described before, we need to define modified ``hatted'' generators
in such a way that the modified, unconstrained generators together with the
anti-ghosts gives a sub-complex $\wh{\Omega}$, i.e. such that the BRST
operator acting on
the unconstrained generators does not involve constrained generators. We find:
\bea
\hat{B}^- & = & B^- - b\G \nn
\hat{B}^+ & = & B^+ + \B c \nn
\hat{U} & = & U - \hf (bc)_0 + \frac{1}{6} (\B\G)_0 \nn
\hat{Y} & = & Y + \thf (\B\G)_0 \nn
\hat{T} & = & T - \hf (\pa b c)_0 - \thf (b \pa c)_0 + \hf (\pa\B\G)_0 +
\thf (\B\pa\G)_0
\ena
while $\hat{G}^+=G^+$ and $\hat{W}^+=W^+$.

the central charge of $\hat{T}$ is $\hat{c} = -\frac{(2k+1)(3k+4)}{k+2}$.
The operator product expansions of the ``hatted'' generators are unchanged,
except for:
\bea
\ope{\hat{T}}{\hat{U}} & = & \ord{3}{-\frac{2}{3}(k+2)} + \ord{2}{\hat{U}}
+ \ordo{\pa\hat{U}} \nn
\ope{\hat{T}}{\hat{Y}} & = & \ord{3}{\frac{2}{3}(k+2)} + \ord{2}{\hat{U}}
+ \ordo{\pa\hat{U}} \nn
\ope{\hat{U}}{\hat{U}} & = & \ord{2}{(3k+8)/18} \nn
\ope{\hat{U}}{\hat{Y}} & = & \ord{2}{-1/9} \nn
\ope{\hat{Y}}{\hat{Y}} & = & \ord{2}{-(3k+4)/18} \nn
\ope{\hat{B}^+}{\hat{B}^-} & = & \ord{2}{-(k+2)} +\ordo{-2\hat{U}-2\hat{Y}}\nn
\ena

We find the generators of the BRST-cohomology, and thereby the generators
of $\cw(s\ell(3|1),s\ell(3))$ to be:
\bea
\ct & = & \hat{T} - 3 \pa \hat{U} \nn
G & = & \hat{G}^+ - (k+2)(\hat{B}\hat{T})_0 - k (\hat{B}^-\pa\hat{U})_0 +
(2k+3) (\hat{B}^-\pa\hat{Y})_0 - 2(k+2) \left[ (\hat{U}\pa\hat{B}^-)_0
-2(\hat{Y}\pa\hat{B}^-)_0
\right]+\nn
& &  + 4 (\hat{B}^-\hat{U}\hat{U})_0 -
4 (\hat{B}^-\hat{U}\hat{Y})_0 + (\hat{B}^-\hat{Y}\hat{Y})_0
 + \hf(k+2)(2k+1) \pa^2\hat{B}^-   \nn
W & = & \hat{W}^+ -k (\hat{B}^-\pa\hat{B}^+)_0
-2(k+2) \left[ (\pa\hat{B}^-\hat{B}^+)_0 - (\hat{T}\hat{U})_0 \right]
+ 2(4+3k) (\hat{U}\pa\hat{U})_0 + 4 (\hat{B}^-\hat{B}^+\hat{U})_0+\nn
& &
- 2(\hat{B}^-\hat{B}^+\hat{Y})_0 - 8 (\hat{U}\hat{U}\hat{U})_0
 + 6 (\hat{Y}\hat{Y}\hat{U})_0 -2(k+2)^2 \pa^2\hat{U} +
2 (\hat{U}\pa\hat{Y})_0 -4 (\hat{Y}\pa \hat{U})_0 \nn
B & = & \hat{B}^+ \nn
\cy & = & \hat{Y}
\ena

Note that except for normal-ordering contributions, we recover the
classical linearization (\ref{lin-cl}).

\subsection{Case of $\cw B_2\sim \cw(osp(1|4),sp(4))$}
\subsubsection{Classical Linearization}

\indent

As a second example, we have chosen the linearization of the algebra
$\cw(osp(1|4), sp(4))$.
The affine $osp(1|4)$ can be parameterized by:
$$
J(z) = \left ( \begin{array}{r|rrrr}
             0 & j_{1}(z)& j_{2}(z)& j_{3}(z)&  j_{4}(z)\\ \hline
      -j_{4}(z)& H_{1}(z)& J_{1}(z)& J_{3}(z)&  J_{4}(z)\\
      -j_{3}(z)& J_{5}(z)& H_{2}(z)& J_{2}(z)&  J_{3}(z)\\
       j_{2}(z)& J_{7}(z)& J_{6}(z)&-H_{2}(z)& -J_{1}(z)\\
       j_{1}(z)& J_{8}(z)& J_{7}(z)&-J_{5}(z)& -H_{1}(z)
\end{array}
\right )
$$
$j_{i}$ denotes the fermionic currents, while $H_i$ and $J_i$ denotes
the bosonic currents. The \cw~algebra $\cw(osp(1|4),sp(1|4))=\cw\cb_2$
is obtained by imposing the constraints
$$
J_c(z) = \left (  \begin{array}{r|rrrr}
             0 &        0&  \psi(z)& j_{3}(z)&  j_{4}(z)\\ \hline
      -j_{4}(z)& H_{1}(z)& J_{1}(z)& J_{3}(z)&  J_{4}(z)\\
      -j_{3}(z)&        1& H_{2}(z)& J_{2}(z)&  J_{3}(z)\\
        \psi(z)&        0&        1&-H_{2}(z)& -J_{1}(z)\\
              0&        0&        0&       -1& -H_{1}(z)
\end{array}
\right )
$$
where $\psi(z)$ is an auxiliary free fermion, normalized such that
$$
\ope{\psi}{\psi} = \ordo{1/2}
$$
We can use the gauge-invariance induced by these first class constraints to
choose the highest weight gauge, which takes the form:
$$
J_{hw}(z) = \left ( \begin{array}{r|rrrr}
       0& 0& 0& 0&  G(z)\\ \hline
      -G(z)& 0&3T(z)& 0&  W(z)\\
       0& 1& 0&4T(z)&  0\\
       0& 0& 1& 0&-3T(z)\\
       0& 0& 0&-1&  0 \end{array} \right )
$$
where $T$ is the energy-momentum tensor, $W$ is a spin 4 primary field, and
$G$ is a spin $\frac{5}{2}$ primary fermionic field.

In order to linearize this algebra, we should consider the algebra
$\cw(osp(1|4),s\ell(2))$, obtained by imposing the constraints and the
highest weight gauge
\be
J_c(z) \ = \ \left ( \begin{array}{r|rrrr}
             0 &        0& j_{2}(z)& j_{3}(z)&  j_{4}(z)\\ \hline
      -j_{4}(z)& H_{1}(z)& J_{1}(z)& J_{3}(z)&  J_{4}(z)\\
      -j_{3}(z)&        1& H_{2}(z)& J_{2}(z)&  J_{3}(z)\\
       j_{2}(z)&        0& J_{6}(z)&-H_{2}(z)& -J_{1}(z)\\
              0&        0&        0&       -1& -H_{1}(z) \end{array} \right )
\ee

\be
J_{hw}(z) \ = \ \left ( \begin{array}{r|rrrr}
             0 &        0& G^{-}(z)&        0&  G^{+}(z)\\ \hline
      -G^{+}(z)&     U(z)&   \tilde{\ct}(z)&        0&  W_{+}(z)\\
              0&        1&     U(z)&        0&         0\\
       G^{-}(z)&        0& W^{-}(z)&    -U(z)&   -\tilde{\ct}(z)\\
              0&        0&        0&       -1&     -U(z) \end{array} \right )
\ee

Using the soldering procedure, we can find the operator product expansions of
this algebra. The normalized energy-momentum tensor is
$\ct=\frac{2}{k}\tilde{\ct}$, with a central charge of $c=-12k$ ($k$ is the
level of the affine $OSp(1|4)$). $U$ is a primary $U(1)$ current, $G^\pm$ are
primary fermionic spin $\frac{3}{2}$ currents with $U(1)$ charge
$\pm \hf$,
and $W^\pm$ are primary bosonic spin 2 currents with $U(1)$ charge $\pm 1$.
The rest of the non-trivial operator product are given {\em in the quantum
form }
in equation (\ref{q-ope}): to get the classical operator product expansions,
one has simply, in each term, to discard all but the leading order in $k$.

In order to find the classical linearization of the $\cw\cb_2$ algebra, we
impose the secondary constraints $\cw(osp(1|4),s\ell(2))$:
$G^-(z)=\psi(z)$ and $W^-(z)=1$. We then make a finite gauge-transformation
$J\rightarrow J^g = gJg^{-1} + k (\prt g)g^{-1}$ where
$g = exp(\Lambda),\,\Lambda\in\cn$, such that the gauge-transformed current
$J^g$ is of the form
\be
\label{J-wb2}
J_(z) = \left ( \begin{array}{r|rrrr}
       0& 0& 0& 0&  G(z)\\ \hline
      -G(z)& 0& \tilde{T}(z)& 0&  W(z)\\
       0& 1& 0&    0&  0\\
       0& 0& 1& 0& -\tilde{T}(z)\\
       0& 0& 0&-1&  0 \end{array} \right )
\ee
Note that while this is indeed the currents of the algebra $\cw\cb_2$, it
is not in the
highest weight gauge, and therefore $G$ and $W$ may not be primary fields in
this basis.

We find that the finite, field dependent gauge-transformation that takes
the constrained current into the gauge (\ref{J-wb2}) is
\bea
\Lambda = \left ( \begin{array}{c|cccc}
             0 &  0 & 0 &     -\psi & \thf \psi U - k \prt\psi \\ \hline
-\thf+k\prt\psi &  0 & -U& U^2-k\prt U&                \epsilon \\
          \psi &  0 & 0 &       -2U &              U^2-k\prt U \\
             0 &  0 & 0 &         0 &                       U \\
             0 &  0 & 0 &         0 &                       0
\end{array} \right ), \\
\epsilon = \psi G^+ - \frac{k^2}{2}\psi\prt^2\psi + 2 \tilde{\ct} U
+ k U\prt U + \frac{k}{6}U\psi\prt\psi - \th U^3 - k^2 \prt^2 U.
\ena
Performing the gauge-transformation, we find the linearization of the algebra:
\bea
\label{cl-lin}
G & = & G^+ + \frac{k}{2} \psi \ct + \frac{k}{2} \psi \prt U + k U \prt \psi
- \hf \psi U^2 - k^2 \prt^2\psi \nn
T & = & \ct - \psi \prt \psi + U^2 - 2 \prt U \\
W & = & W^+ - 3 k G^+ \prt \psi + k \psi \prt G^+ + 2 G^+ \psi U
+ \ct \prt U + \frac{k^2}{2} U \prt \ct  - \frac{k}{2} \ct U^2
- \frac{k^2}{2} \ct \psi \prt \psi +\nn
& & - \frac{k^3}{2}( \psi \prt^3\psi +3 \prt \psi \prt^2\psi )
+  \frac{k^2}{2} U \prt^2U - \frac{k^2}{4} \prt U \prt U
- \frac{k^3}{2} \prt^3U
 + k^2 (\psi U \prt^2\psi
-  \psi \prt \psi \prt U )\nonumber
\ena

\subsubsection{Quantum Linearization}

\indent

To perform the quantum linearization, we perform first the
quantum hamiltonian reduction leading to $\cw(osp(1|4),s\ell(2))$. As a result
of this procedure, we get the generators of  the algebra in terms of the
generators of $OSp(1|4)$ (which we do not need for the linearization), and
the operator product expansions. We find the central charge to be
$c=-\frac{(4k+7)(6k+5)}{2k+5}$. $U$ is a primary $U(1)$ current, $G^\pm$ are
primary fermionic spin $\frac{3}{2}$ currents with $U(1)$ charge $\pm \hf$,
and $W^\pm$ are primary bosonic spin 2 currents with $U(1)$ charge $\pm 1$.
The rest of the non-trivial operator product expansions are:
\bea
\label{q-ope}
\ope{U}{U} & = & \ord{2}{k+\frac{7}{4}} \nn
\ope{G^\pm}{G^\pm} & = & \ordo{\pm \hf W^\pm } \nn
\ope{G^+}{G^-} & = & \ord{3}{-(1+k)(7+4k)/4} + \ord{2}{-\hf(k+1) U}
+ \ordo{-\ei(2k+5)T - \fo (UU)_0 - \fo(k+1) \prt U} \nn
\ope{G^\pm}{W^\mp} & = & \ord{2}{\mp \hf (3k+5) G^\mp}
 +\ordo{\mp \hf(k+2) \prt G^\mp - (UG^\mp)_0}      \nn
\ope{W^+}{W^-} & = & \ord{4}{(k+1)(3k+5)(4k+7)} + \ord{3}{(k+1)(3k+5) U} +\nn
& & + \ord{2}{-\fo(2k+3)(2k+5)T + \hf(5+4k) (UU)_0
+ \hf(k+1)(3k+5) \prt U} +\nn
& &  + \ordo{-\hf(2k+5) (TU)_0 + (UUU)_0 - 2(G^-G^+)_0
- \fo(k+1)(2k+5) \prt T }+\nn
& &  + \ordo{2(k+1) (U\prt U)_0 +\fo(2k^2+8k+9) \prt^2 U}
\ena

In order to perform the secondary quantum hamiltonian reduction, we now need
to impose the secondary constraints $G^-(z) = \psi(z)$ and $W^-(z)=1$. In
order to do this we introduce a ghost pair corresponding to each constraint,
a fermionic pair $(b,c)$ for the constraint $W^-(z)=1$, and a bosonic pair
$(\beta,\gamma)$ (with OPE $\ope{\gamma}{\beta}=\ordo{1}$)
for the constraint $G^-(z) = \psi(z)$.
The BRST current then takes the form
$$
j = (W^--1)c + (G^--\psi)\gamma + \fo b\gamma\gamma
$$
Now, we have to introduce the ``hatted'' generators that are the starting
point of the tic-tac-toe
construction (see section \ref{linqa}). They
are defined by
\bea
\hat{G}^- & = & s(\beta) + \psi \quad = \quad G^- + \hf \gamma b \nn
\hat{W}^- & = & s(b) + 1        \quad = \quad W^-  \nn
\hat{U} & = & U - \fo (\beta\gamma)_0 + \hf (bc)_0 \nn
\hat{T} & = & T - 2 (b\prt c)_0 - (\prt bc)_0 - \hf (\gamma\prt\beta)_0 -
\thf (\prt\gamma\beta)_0   \nn
\hat{G}^+ & = & = G^+ + \hf(3k+5)\beta\prt c + (k+2) \prt\beta c - U\beta c
- \fo (\beta\gamma\beta c)_0   \nn
\hat{W}^+ & = & W^+ + 2 G^+\beta c + \frac{3k+5}{2} \beta\beta\prt cc
\ena
The generators that will be used for the linearization are
$\hat{U}$, $\hat{G}^+$ and $\hat{T}$, together with the free fermion $\psi$.
The modified central charge is $\hat{c} = -\frac{2(12k^2+46k+55)}{2k+5}$, and
the operator product expansions of the ``hatted'' generators are:
\bea
\ope{\hat{T}}{\hat{U}} & = & \ord{3}{-2} + \ord{2}{\hat{U}}
+ \ordo{\prt\hat{U}} \nn
\ope{\hat{U}}{\hat{U}} & = & \ord{2}{k+\frac{5}{2}} \nn
\ope{\psi}{\psi} & = & \ordo{1/2},
\ena
while the rest of the operator product expansions are unchanged. We now find
that the generators of the zeroth cohomology of $s$, i.e. the generators
of $\cw\cb_2$, are:
\bea
T & = & \hat{T} - 2 \prt \hat{U} - (\psi\prt\psi)_0 \nn
G & = & \hat{G}^+ + \fo(2k+5) \psi\hat{T} + \hf(k+1) \psi\prt\hat{U}
+ (k+2) \prt\psi\hat{U} - \hf(\psi\hat{U}\hat{U})_0 - (2+k)^2 \prt^2\psi \nn
W & = & \hat{W}^+ - (7+3k) \hat{G}^+\prt\psi
- (1+k) \prt\hat{G}^+\psi
- \frac{(5+2k)(11+6k)(17+8k)}{192} (\psi\prt^3\psi)_0 +\nn
& & + \hf(2+k)(2k+5) (\hat{T}\prt\hat{U})_0
+ \fo(9+8k+2k^2) (\hat{U}\prt^2\hat{U})_0 +\nn
& & - \frac{3(5+2k)(69+66k+16k^2)}{64} (\prt\psi\prt^2\psi)_0
- \fo(2+k)(6+k) (\prt\hat{U}\prt\hat{U})_0 + 2(\hat{G}^+\psi\hat{U})_0 +\nn
& & - \frac{(5+2k)^2}{8} (\hat{T}\psi\prt\psi)_0
+ \hf(k+2)(2k+5)(\hat{U}\psi\prt^2\psi)_0 + \fo (k+2)(2k+5)
(\prt\hat{T}\hat{U})_0+\nn
& & - \fo(3+2k)(5+2k) (\prt\hat{U}\psi\prt\psi)_0
- \fo (2k+5) (\hat{T}\hat{U}\hat{U})_0
- \hf(2k+3) (\hat{U}\hat{U}\prt\hat{U})_0 +\nn
& &  + \fo (\hat{U}\hat{U}\hat{U}\hat{U})_0
- \frac{(2+k)(73+58k+12k^2)}{24} \prt^3\hat{U}
\ena
Thus, we find in a very simple and natural way the quantum linearization
of $\cw\cb_2$, as it was
given by brute force in \cite{BeKrSo}. We
notice that if in each term we keep only the highest order
in $k$, this expression becomes
the classical linearization
(\ref{cl-lin}).

\sect{Conclusion}

\indent

In this paper, we have presented a general framework and explicit realization
for the linearization of \cw-superalgebras. This
linearization relies on the concept of secondary reductions, that is the
Hamiltonian reduction of
\cw-algebras themselves. The techniques we use ensures that the linear
algebras we obtain have only
fields of positive spin. The price to pay is that some \cw-(super)algebras
 are not linearizable through our procedure. For some of them
(as superconformal algebras)
 we already know that they are linearizable with fields of negative
spins: it should be interesting to
 see whether it is a general feature, or, on the contrary, if there
are other schemes of linearization that
 use only positive spins.

\indent

When considering \cw-superalgebras that contain a true $N=1$ supersymmetric
subalgebra
(Ramon-Neveu-Schwarz superconformal algebra), one can directly perform the
Hamiltonian
reduction is $N=1$ superfield formalism. In that case, one considers
$osp(1|2)$ subalgebras
instead of $s\ell(2)$ embeddings. This techniques applies also to the
secondary reductions, and
therefore to the linearizations. We have not studied exhaustively this
approach, but as the gradations
one uses in $N=1$ formalism are the same as ours, one can already conclude that
this formalism does not provide new schemes of linearizations. In
particular, the $\cw[s\ell(m|m\pm1), s\ell(m|m\pm1)]$ are still not
linearizable in super-fields
formalism, although they are supersymmetric. The same thing appears
for $osp(2m\pm1|2m)$,
$osp(2m|2m)$, and $osp(2m+2|2m)$ algebras.

\indent

Finally, let us mention that the linearized $\cw_3$ algebra has been
used to build non-critical
$\cw_3$ BRST operators as well as new realization of the $\cw_3$ algebra
\cite{LPSX}: such an
approach using the general framework of secondary reductions could indeed
lead to a wide class
of new realizations of \cw-(super)algebras and also to their non-critical
BRST operators.

\indent

\noindent
{\large{\bf Acknowledgments}}

\indent

We would like to thank Jan de Boer and Tjark Tjin for stimulating discussion.
One of the authors (JOM) wishes to thank the Danish Natural Science Research
Council for
financial support.

\appendix
\sect{Construction of the algebra sub-complex $\wh{\Omega}$}
\subsection{$\wh{\Omega}$ is a sub-complex.}

In this section, we show that we can define modified (``hatted'') generators
corresponding to the unconstrained currents, in such a way that the space
$\wh{\Omega}$ generated by these hatted generators and the $c$'s is a
subcomplex. We will do this by a double induction, using the conformal
dimension and the ($H-H'$)-grade of the generators as induction parameters.

We consider the ``twisted'' algebra, i.e. the algebra where the conformal
dimensions are given by the $H'$-grade + 1. In this case the conformal
dimensions of all the constrained generators is 1 and the
($H-H'$)-grade of the constrained generators is less than zero.

We will need a lemma:
\begin{lem}
Take an unconstrained generator $W^{\bar{\A}}$ with conformal dimension
$h$ and grade $n$, and consider $s(W^{\bar{\A}})$. All
generators in this
expression are the result of OPEs between a constrained generator
in $j_{brs}$, and $W^{\bar{\A}}$. Thus all monomials\footnote{We use the word
``monomial'', even though what we have is actually a normal-ordered product.}
of generators
occurring in $s(W^{\bar{\A}})$ must have conformal dimension $h$ and grade
less than $n$. Write:
$$
s(W^{\bar{\A}}) = P_{\bar{\B}}(c)W^{\bar{\B}} +
Q_{\A\bar{\G}}(c) W^\A W^{\bar{\G}} + \cdots,
$$
then we see that the conformal dimension of $W^{\bar{\B}}$ is $h$ and the
grade is less than $n$. The conformal dimension of $W^{\bar{\G}}$ is $h-1$
etc., i.e. all unconstrained generators occurring in $s(W^{\bar{\A}})$ has
{\em either} conformal dimension less than $h$ {\em or} conformal dimension
$h$ and grade less than $n$.
\end{lem}

Assume that we have already found hatted generators for all generators with
conformal dimension less than $h$,
and define $\wh{\Omega}^{h-1}$ to be the space generated
by these hatted generators and the $c$'s.
Assume that $W^{\bar{\A}}$ is any generator with conformal dimension $h$ and
grade 0, we will show that we can define $\wh{W}^{\bar{\A}}$ such
that $s(\wh{W}^{\bar{\A}}) \in \wh{\Omega}^{h-1}$.
Consider $s(W^{\bar{\A}})$.
According to the lemma, all unconstrained generators
occurring in $s(W^{\bar{\A}})$ must have conformal dimension less than $h$.
We can therefore write
$$
s(W^{\bar{\A}}) = \sum_{i,j} A_{ij} B_j,\quad A_{ij} \in \cb,\,
B_j \in \wh{\Omega}^{h-1}
$$
where the $B_j$'s are chosen to be linearly independent. Since $j_{brs}$ is
linear in
the constrained currents,
each of the terms $A_{ij}$ are monomials in the constrained currents, the
$W^\A$'s. Let us consider only those terms that have the highest grade,
considered as monomials in $W^\A$.
$$
s(W^{\bar{\A}}) = \sum_{i,j} A^m_{ij} B_j + \mb{lower orders terms},\quad
A^m_{ij} \mbox{ is order $m$ in $W^\A$}
$$
Now apply $s$ once again. We get:
$$
0 = \sum_{i,j} \left ( s(A^m_{ij}) B_j \pm A^m_{ij} s(B_j) \right )
$$
We know that $s(B_j) \in \wh{\Omega}^{h-1}$, and
$s(A^m_{ij}) \in \cb$. We also know that $s(A^m_{ij})$ is of order
$m+1$ in the $W^\A$'s, and these are the only possible terms of order $m+1$;
and since the expression must vanish order by order in the $W^\A$'s, we find
$$
0 = \sum_{i,j} s(A^m_{ij}) B_j
$$
Since the $B_j$'s are linearly independent we find that
$$
0 = \sum_i s(A^m_{ij})
$$
Now we use the fact that $\cb$ has trivial cohomology. Since
$\sum_i A^m_{ij}$ is in the kernel of $s$ it must be in the image of $s$, so
we can find $X_j$ (of grade $m-1$ in the $W^\A$'s)
such that $s(X_j) = \sum_i A^m_{ij}$. Define
$$
W_{(1)} = W - \sum_j X_j B_j.
$$
We find that:
\beano
s(W_{(1)}) & = &  \sum_{i,j} A^m_{ij} B_j + \mb{lower orders terms}
- \sum_j ( s(X_j) B_j \pm  X_j s(B_j) ) \\
& = & \sum_{i,j} A^m_{ij} B_j + \mb{lower order terms}
- \sum_{i,j} A^m_{ij} B_j
\mp \sum_{j} X_j s(B_j) \\
& = & \mbox{lower order terms } \mp \sum_{j} X_j s(B_j)
\enano
(the $\pm$ depends on the Grassman parity of $X_j$).
All these terms are of order at most $m-1$ in the $W^\A$'s. By induction we
see that we can define $\hat{W}^{\bar{\A}}$ such that $s(\hat{W}^{\bar{\A}})$
is a polynomial of
degree 0 in the constrained currents.

We want to show that in fact no $b$'s appear in $s(\hat{W}^{\bar{\A}})$
either.
Actually this is quite simple: write
$$
s(\hat{W}^{\bar{\A}}) = B + \sum_\A B_\A b^\A + \sum_{\A,\B} B_{\A\B}
b^\A b^\B
+ \cdots.
$$
Apply $s$ again to get
$$
0 = s(B) + \sum_\A s(B_\A) b^\A \pm B_\A (\hat{J^\A}-\chi^\A) + \cdots
$$
Since $s(B_\A)$ does not contain any constrained currents, we must have
$0 = \sum_\A B_\A \hat{J}^\A$, but this can only be true if $B_\A=0$ for all
$\A$.
We see that indeed $s(\hat{W}^{\bar{\A}}) \in \wh{\Omega}$.

Now assume that we have found hatted generators for all generators with
conformal dimension less than $h$, and with conformal dimension $h$ and
grade less than $n$,
and define $\wh{\Omega}^{h}_{n-1}$ to be the space generated
by these hatted generators and the $c$'s.
Assume that $W^{\bar{\A}}$ is any generator with conformal dimension $h$
and grade $n$, we want to show that we can define $\wh{W}^{\bar{\A}}$ such
that $s(\wh{W}^{\bar{\A}}) \in \wh{\Omega}^{h}_{n-1}$.
Consider $s(W^{\bar{\A}})$.
According to the lemma,
any unconstrained generator $W^{\bar{\B}}$ that occurs in
$s(W^{\bar{\A}})$ has {\em either} have conformal dimension less than $h$
{\em or} conformal dimension $h$ and grade less than $n$.
We can therefore write
$$
s(W^{\bar{\A}}) = \sum_{i,j} A_{ij} B_j,\quad A_{ij} \in \cb,\,
B_j \in \wh{\Omega}^{h}_{n-1}.
$$
We can therefore repeat the arguments from above word by word to define
$\hat{W}^{\bar{\A}}$ such that $s(\hat{W}^{\bar{\A}}) \in \wh{\Omega}$.

We have shown that to any generator $W^{\bar{\A}}$ we can construct
$\hat{W}^{\bar{\A}}$ such that $s(\hat{W}^{\bar{\A}}) \in \wh{\Omega}$.
We have therefore shown that $\wh{\Omega}$ is a sub-complex.

\subsection{Construction of the algebra law in $\wh{\Omega}$}

In general, the subcomplex $\wh{\Omega}$ generated by
$\{\hat{W}^{\bar{\A}},c_\A \}$ is not {\it a priori} a subalgebra.
Actually it turns out that $\wh{\Omega}$ is ``often'' a subalgebra in
explicit examples, but we can find cases where this is not the case.
It turns out that if the OPEs between {\em constrained} and
{\em un-constrained} operators contains terms that are multi-linear in
the {\em constrained} generators, then extra operators (in the simplest cases
of the form $(bc)_0$) will appear in the OPEs of the generators of
$\wh{\Omega}$.
We will argue that even in this case, it is consistent to project the
OPEs on $\wh{\Omega}$, thereby making $\wh{\Omega}$ an algebra.

Define $\ca$ to be the ghost-number zero subspace of $\wh{\Omega}$, i.e. the
space generated by the $\{ \hat{W}^{\bar{\A}} \} $, and let
$a,b \in \ca$. Then, we have $a\circ b$ in $\Omega$, where $\circ$ is
the algebra composition (the OPEs) in $\Omega$.
We define a new composition on $\wh{\Omega}$ through
\be
a\diamond b=\pi(a\circ b), \mbox{ where } \pi\mbox{ is the projection on }\ca
\ee

The generators of the \cw-algebra is constructed by the tic-tac-toe
construction as polynomials in the generators of $\wh{\Omega}$ with quantum
number zero, i.e. the generators of the \cw-algebra are polynomials of
the generators in $\ca$. Since the algebra is closed, we know that for any
\cw-generators $V$ and $W$, in
$V\circ W$ appears only polynomial in the generators of $\ca$.
Thus from the point of view of the \cw-algebra, one can consistently
do the projection $\pi$. This does not change the OPEs of the \cw-algebra.

It does not immediately follow, however, that the ``$\diamond$'' OPEs
gives an associative algebra.
We have investigated two explicit examples where the $\circ$ does not give
an algebra on $\wh{\Omega}$ (the linearization of $\cw_5$ and $\cw_6$; for
higher $n$ the calculations becomes extremely time consuming even on the
computer), and in these cases the algebra is indeed associative. We expect
this to be the case in general.

Note that the problem is inherent in the method that we use for the
hamiltonian reduction, the BRST method; it is {\em not} connected
directly to the quantization. Indeed, if we perform the {\em classical}
hamiltonian reduction using the classical BRST method (see e.g. \cite{who}),
we find that also in that case
the classical OPEs of the generators of $\wh{\Omega}$ contains extra
operators, in the simplest examples of the form $bc$.

On the other hand, we can use the alternative quantization approach that has
already been mentioned earlier. In this approach, we start from
the classical \cw-algebra $\cw(\cg,\ch')$ and find the classical expressions
for the generators in $\cw(\cg,\ch)$ in terms of the unconstrained
generators of $\cw(\cg,\ch')$. We adjust the various
coefficients (in the expressions for the generators of $\cw(\cg,\ch)$ and in
the expressions for the OPEs of the unconstrained generators of
$\cw(\cg,\ch')$), in such a way that the
quantum OPEs of the generators of $\cw(\cg,\ch)$ gives a closed algebra.
In that case
clearly no extra ghost-terms appear, and it is clear that we do not need to
modify the OPEs; everything will work as in the classical case.
It seems evident that the quantum OPEs of the un-constrained generators of
$\cw(\cg,\ch')$ obtained in this way are
identical to the OPEs defined by the $\diamond$.

In particular, if we focus on the classical hamiltonian reduction, it is
clear that the gauge approach described in section \ref{truc} will provide
a ``good'' linearization, while the classical BRST approach already leads to
the emergence of $bc$-type terms. In that case it is obvious that the
``$\diamond$'' composition law will just reproduce the classical Poisson
brackets obtained by the gauge-method.

\end{document}